\documentclass[reprint,superscriptaddress,preprintnumbers,nofootinbib,nobibnotes,amsmath,amssymb,amsfonts,aps,prd,floatfix]{revtex4-2}

\usepackage{amsbsy,esint,mathrsfs,commath,amsfonts,mathtools,amsthm}
\usepackage[usenames,dvipsnames]{xcolor}
\usepackage{physics}
\usepackage{graphicx} 
\usepackage{hyperref}
\usepackage[caption=false]{subfig}

\hypersetup{
  colorlinks=true,
  linkcolor=Blue,
  filecolor=magenta,
  urlcolor=Blue,
  citecolor=Blue,
}

\def\closurec{\vec{\mathcal{X}}}

\def\Hfull{\breve H}
\newcommand{\Hfullred}{\Hfull_{\mbox{\tiny red}}}

\def\af{{\mathcal{F}}}

\newcommand{\ed}{\text{d}}

\newcommand{\bmath}[1]{\mbox{\boldmath $#1$}}
\def\Lie{\mathcal{L}}
\def\equalc{\stackrel{\mathcal{C}}{=}}

\newcommand{\cc}[1]{\,\overline{\!{#1}\!}} 

\newcommand{\boxbra}[1]{[ #1 |}
\renewcommand{\ket}[1]{| #1 \ra}
\renewcommand{\bra}[1]{\la #1 |}
\renewcommand{\braket}[2]{\la #1 | #2 \ra}
\newcommand{\la}{\langle}
\newcommand{\ra}{\rangle}
\newcommand{\boxket}[1]{| #1 ]}

\newcommand{\U}{\text{U}}
\newcommand{\SU}{\text{SU}}
\newcommand{\alsu}{\mathfrak{su}}

\def\lapseM{\mathcal{{M}}}
\def\lapseN{\mathcal{N}}
\def\lapseRW{\tilde{\lapseN}}

\def\trw{\tau}

\def\onee{\omega}

\def\gammam{h}
\def\volg{V_h}
\def\scalef{\tilde{a}}

\def\expansio{\vartheta}
\def\expansion{\theta}
\def\expansionrw{\Theta}

\newcounter{mnotecount}
\newcommand{\mnote}[1]
{\protect{\stepcounter{mnotecount}}$^{\mbox{\footnotesize $\bullet$\themnotecount}}$ 
\marginpar{
\raggedright\tiny
$\!\!\!\!\!\!\,\bullet$\themnotecount: #1} }

\begin{document}
 
\title{Homothetic expansion of polyhedra in the two-vertex model: emergence of FLRW}

 \date{\today}

 \author{I\~naki Garay} 
 \email{inaki.garay@ehu.eus}
 \affiliation{Department of Physics and EHU Quantum Center,
   University of the Basque Country UPV/EHU, Barrio Sarriena s/n, 48940, Leioa, Spain}  
 \author{Sergio Rodr\'iguez-Gonz\'alez}
 \email{sergio.rodriguez@ehu.eus}
 \affiliation{Department of Physics and EHU Quantum Center,
   University of the Basque Country UPV/EHU, Barrio Sarriena s/n, 48940, Leioa, Spain} 
 \author{Ra\"ul Vera} 
 \email{raul.vera@ehu.eus}
 \affiliation{Department of Physics and EHU Quantum Center,
   University of the Basque Country UPV/EHU, Barrio Sarriena s/n, 48940, Leioa, Spain}  
\begin{abstract}
  The cosmological behavior associated to a $\U(N)$-symmetry reduced sector of the  loop-quantum-gravity truncation known as the two-vertex model is further explored in this work. We construct convenient frame bases
  that encode the whole classical phase space of the twisted geometry
  associated to the graph.
  We show that the polyhedra of the twisted geometry suffer under evolution an homothetic expansion,
  which strengthens the correspondence to the Robertson-Walker geometry.
\end{abstract}

\maketitle

\tableofcontents

\section{Introduction}
Loop quantum gravity (LQG) is a quantum theory of gravity whose Hilbert space is constituted by spin network states, i.e., $\SU(2)$ wave functions defined over oriented graphs with spins attached to the links and intertwiners to the vertices \cite{rovellibook}.

The study of the truncated Hilbert space defined by a fixed graph has proved to be an interesting arena to study the physical effects of the loop quantization. Such a truncation is described by the spin network states with support on that graph, obtained from the quantization of the holonomy-flux phase space on the graph.
The amount of data contained in a truncated model is finite, and thus captures only a finite number of degrees of freedom. In that sense, a fixed graph provides a first approximation (discretization) to the geometry \cite{rovellibook,rovelliGeometryLoopQuantum2010}. 

The spin networks are eigenstates of the area and volume operators. Therefore, they represent quanta of area and volume associated to the links and nodes respectively \cite{Rovelli:1994ge, Ashtekar_Lewandowski_1997}.
At the classical level, the framework given by the spinorial formalism for LQG  provides a geometrical interpretation for this notion of discretization in terms of convex polyhedra attached to the nodes of the graph \cite{Livine2011,Livine:2013zha,DupuisSpinorsAndTwistors}. These polyhedra would constitute the classical analogue of the quanta of volume predicted by LQG, so that they would provide a discretization of classical physical space known as \textit{twisted geometries}, which is as a generalization of Regge geometries where the shape matching conditions are relaxed \cite{DupuisSpinorsAndTwistors,freidelTwistedGeometriesGeometric2010a, freidelTwistorsTwistedGeometries2010}.

The application of the spinorial formalism of LQG \cite{Borja_Freidel_Garay_Livine_2010_Return,Livine2011,Livine:2013zha,DupuisSpinorsAndTwistors} to truncated models (in particular to the two-vertex model, given by two nodes and an arbitrary number of links $N$), has led to numerous analytical and numerical results describing the time evolution of the discrete classical geometry of polyhedra. In particular, those results have allowed to identify reduced sectors within the model that mimic cosmological scenarios, thus pointing out a correspondence between the classical phase space of twisted geometries and general relativity  \cite{Borja_Freidel_Garay_Livine_2010_Return, Borja_Diaz-Polo_Garay_Livine_2010_Dynamics,Livine_Martin-Benito_ClassicalSettingEffective2013, Aranguren_Garay_Livine_2022, Cendal:2024uzu,Diego_Garays_2025}. This last point was extensively developed in \cite{Cendal:2024uzu}, where it was shown that the dynamical equations of the $\U(N)$-reduced sector of the model reproduced the Friedmann equation in the low curvature regime, as well as the old dynamics of loop quantum cosmology (LQC). Moreover, it was suggested that polyhedra could be undergoing an isotropic (homothetic)  expansion, which could be indicative of an emergent Robertson-Walker (RW) geometry.
  It is in this sense, based on the combination of these two facts, that it has been argued that the  $\U(N)$-reduced sector of the two-vertex model recovers the Friedmann-Lema\^itre-Robertson-Walker (FLRW) cosmological model in low curvature regimes.\footnote{We adopt the convention of referring to the FLRW model as the RW geometry plus the Einstein field equations. Observe that the quantum corrections coming from LQC only modify the field equations. }

In this paper, we further explore the dynamical properties of the two-vertex model and the cosmological behavior associated with its $\U(N)$-reduced sector. 
After a short review of the spinorial and twisted geometry formalisms in section \ref{formalisms},
in section \ref{Sec_Frame_basis} we define the frame bases, and show that they encode all the twisted geometries information on a graph,  giving a full description of the dynamics on the two-vertex model. These results are used in section \ref{sec:unsector} to prove that, in the $\U(N)$-reduced sector of the model, polyhedra undergo an homothetic expansion,  as it was suggested in \cite{Cendal:2024uzu}.
This fact, together with the results of \cite{Cendal:2024uzu}, is finally used in section \ref{Sec_cosmo} to stablish a complete correspondence between the geometry of the $\U(N)$-reduced sector of the two-vertex model given by the twisted geometries, and the RW geometry.

\section{Spinor formalism and twisted geometries}\label{formalisms}

The classical description of the holonomy-flux phase space used in LQG is 
represented  by a copy of a
pair $(g,X)\in \SU(2)\times \alsu(2)$ associated to each link of the graph
\cite{freidelTwistedGeometriesGeometric2010a,
  freidelTwistorsTwistedGeometries2010,DupuisSpinorsAndTwistors,
  Livine2011,Livine:2013zha},
where $g$ is the holonomy at the link and 
$X:=\vec{X}\cdot \vec{\sigma}=: X^I \sigma_I$ the flux,
with $\vec{X}\in\mathbb{R}^3$, $\vec{\sigma} = (\sigma^1,\sigma^2,\sigma^3)$
the Pauli matrices, and we use $I,J,\ldots=1,2,3$ for indices in $\mathbb{R}^3$.
If we use $A,B,\ldots=0,1$ for the components in the defining representation, the
Poisson algebra satisfied by $g$ and $X$ is given by \cite{Livine2011,Livine:2013zha}
 \begin{align}
& \left\{g_{AB},g_{CD}\right\} = 0,  \label{HolonomyFluxAlgebra1}\\
 & \left\{X^I ,X^J \right\} = {\epsilon}{^{IJK}}X^{K},\quad
    \left\{\vec{X},g_{AB}\right\} = -i\frac{1}{2}g\vec{\sigma}_{AB},
    \label{HolonomyFluxAlgebra2}
\end{align}
where $\epsilon^{IJK}$ denotes the Levi-Civita symbol.

The spinorial formalism for LQG \cite{Livine2011,Livine:2013zha,
Borja_Freidel_Garay_Livine_2010_Return,DupuisSpinorsAndTwistors}
offers a suitable parametrization of this phase space using
elements of $\mathbb{C}^2$, that is, spinors.
Expressed in components $z^A=\{z^0,z^1\}\in \mathbb{C}$, the
spinors $\ket{z}$ and their conjugates $\bra{z}$ are written as
\begin{equation}
  \ket{z} = \begin{pmatrix}
    z^{0}\\
    z^{1}
  \end{pmatrix},\quad
  \bra{z}=\left(\cc{z^0},\cc{z^1}\right),
\end{equation}
while the corresponding dual $|z]$, satisfying $\bra{z}z]=0$ and
$[z|z]=\braket{z}{z}$,  is defined by 
\begin{equation}
  |z] = 
  \begin{pmatrix}
    - \cc{z^{1}}\\
    \cc{z^0}
  \end{pmatrix}.
\end{equation}

In order to write the holonomies and the fluxes in terms of the
spinors, one considers two spinors, $\ket{z^s}$ and $\ket{z^t}$,
associated to the source and target nodes of the link respectively\footnotetext{
In what follows we extend the use of superindexes $s$ and $t$ to
label objects relative to the source or target nodes respectively.}.
With this notation, the holonomy of the link reads \cite{Livine2011}
\begin{equation}
  g = \frac{|z^t] \bra{z^s} - \ket{z^t} [z^s|}{\sqrt{\braket{z^t}{z^t} \braket{z^s}{z^s}}},\label{DEF holonomia}
\end{equation}
while the flux is given by either one of the two vectors associated to the source or target nodes,
\begin{equation}
  \Vec{X}^{s,t}:= \frac{1}{2}\bra{z^{s,t}} \vec{\sigma} \ket{z^{s,t}}=-\frac{1}{2}\boxbra{z^{s,t}}\Vec{\sigma}\boxket{z^{s,t}}.\label{DEF vectores normales}
\end{equation}
These two vectors are linked by the relation
\begin{equation}
\frac{X^t }{|\vec{X}^t|} =  - g \frac{X^s}{|\Vec{X}^s|} g^{-1}\label{relation fluxes}
\end{equation}
in $ \alsu(2)$ by construction, where
$|\Vec{X}^{(s,t)}|= \frac{1}{2}\la z^{(s,t)}|z^{(s,t)}\ra$ is the norm.

At this point, by equipping
$\mathbb{C}^2\times\mathbb{C}^2$ with the Poisson brackets
\begin{equation*}
    \left\{z^{sA},\cc{z^{sB}}\right\} =  
    \left\{z^{tA},\cc{z^{tB}}\right\}=-i\delta^{AB},
  \end{equation*}
  while keeping the rest zero, the brackets \eqref{HolonomyFluxAlgebra2} are recovered.
    Regarding  \eqref{HolonomyFluxAlgebra1}, a straigforward calculation leads to
    \begin{align*}
     & \{g_{AB},g_{CD}\}= \frac{\braket{z^s}{z^s}- \braket{z^t}{z^t}}{\braket{z^t}{z^t} \braket{z^s}{z^s}}\\
                    &\quad    \times \left(   z^t_A \cc{z^s_B }\, \epsilon^{ }_{CE} \, \epsilon^{ }_{DF} 
                          -   z^t_C \cc{z^s_D }\, \epsilon^{ }_{AE} \epsilon^{ }_{BF} \right) \cc{z^{tE} }z^{sF}.
      \end{align*}
    Therefore, to regain \eqref{HolonomyFluxAlgebra1}, and thus the whole algebra satisfied by $(g,X)$,
   it is necessary and sufficient that  the so-called matching constraint at the link
\begin{equation}
    \mathcal{C} := 2(|\vec{X}^s| - |\vec{X}^t|) = 
    \braket{z^s}{z^s}-\braket{z^t}{z^t}=0 
    \label{DEF matching constraint}
\end{equation}
is imposed. This reflects the fact that in LQG there is a unique irreducible representation of $\SU(2)$ associated to each link. The transformation on the spinors generated by
the matching constraint is a $\U(1)$ transformation of the form
\begin{align}
  (\ket{z^s},\ket{z^t}) \to (e^{i\phi}\ket{z^s},e^{-i\phi}\ket{z^t}),\quad \phi \in \mathbb{R},\label{U1}
\end{align}
that does not modify the pair $(g,X)$.

The above parametrization is extended to all the graph by taking the cartesian product
of the phase spaces at each link of the graph and extending the algebra by setting the
Poisson brackets between quantities  of different links to vanish.
Specifically, given a graph consisting of $N$ links, if we label by latin subindices $i,j... =1,\ldots, N$
the quantities associated to the links,
the algebra of the spinor representation is given by
\begin{equation}
  \left\{z_i^{sA},\cc{z_j^{sB}}\right\} =  \left\{z_i^{tA},\cc{z_j^{tB}}\right\}=-i\delta^{AB}\delta_{ij}
  \label{poissonbrackets}
\end{equation}
and zero the rest.

The $\SU(2)$ invariance of the LQG intertwiners
at the nodes is recovered in the spinorial formalism
by imposing the so-called closure constraints at each node $\nu$ of the graph
\begin{equation}
  \closurec^\nu := \sum_{i\in \nu}\vec{X}^\nu_{i} = 0,\label{DEF closure constraint}
\end{equation}
where 
the sum is extended to all the links $i$ joining the node $\nu$.

The area and volume operators constructed within the LQG theory 
allow us to interpret the spin networks as representations of 
quanta of area associated to the links (dependent on the spin label) and
chunks of volume attached to the nodes (that depend on the intertwiner). 
In this sense, a truncation to a fixed graph  provides a first 
approximation (discretization) to the geometry
\cite{rovellibook,rovelliGeometryLoopQuantum2010}. 
Although the proper realization of this idea arises after
quantization, at the classical level the $\mathbb{R}^3$ vector representation
of the spinorial formalism provides a geometrical interpretation 
for this notion of discretization.

Indeed,  the application of Minkowski's theorem \cite{aleksandrov_convex_polyhedra} to the closure
constraint \eqref{DEF closure constraint} at each node
allows the association of the vectors $\vec X_i^\nu$ at each node
$\nu$ to faces  of area $|\vec X_i^\nu|$ orthogonal to them,
forming a convex polyhedron.
In turn, the matching constraint
\eqref{DEF matching constraint} at each link
ensures that the area of the faces associated to 
the source and target vectors (their norms) are equal.

This representation provides a more general (and abstract) setting than the Regge
geometries (see e.g. \cite{rovelliGeometryLoopQuantum2010})
given that, in this case, the number of edges of the faces at each node and their shapes are not fixed.
On the other hand, given the matching constraint, the information
carried by the pair of vectors $\vec X^s,\vec X^t$ at each link
needs to be complemented
by an extra quantity to encode the six degrees of freedom per link associated to the pair
$(g,X)$. That extra quantity is the twist angle $\xi$, which is to hold
information relative to the extrinsic geometry \cite{freidelTwistedGeometriesGeometric2010a,
freidelTwistorsTwistedGeometries2010}.
The set of variables
$(\hat{X}^s,\hat{X}^t,A,\xi)$, where the hat stands for the normalized
vector and $A=|\vec X^s|=|\vec X^t|$, at each link, plus the closure constraint
at each node, determine the
twisted geometry framework. In \cite{freidelTwistedGeometriesGeometric2010a,
freidelTwistorsTwistedGeometries2010} the explicit canonical transformation
between the holonomy-flux pair $(g,X)$ and the set $(\hat{X}^s,\hat{X}^t,A,\xi)$
at each link is constructed.

\section{Frame basis for the twisted geometries}
\label{Sec_Frame_basis}

\begin{figure}
    \centering
    \includegraphics[width=0.4\textwidth]{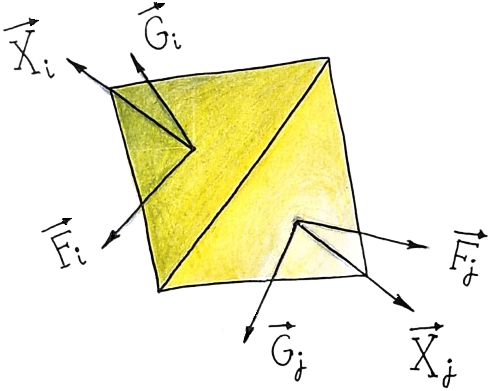}
    \centering
    \caption{Representation of the vectors of the frame bases attached to the faces $i$ and $j$ of a polyhedron. Vectors $\vec{X}$ are normal to the  faces (pointing outwards) and $\vec{G}$, $\vec{F}$ are tangent to each face (with $\vec{G}\cdot\vec{F}=0$).   
  }
    \label{figure_frame_basis}
\end{figure}

Following the scheme described in the previous section, given a polyhedron and the set of spinors $\{\ket{z_{i}}\}$ attached to its faces, we define the associated
\textit{generalized normal vectors} and \textit{generalized tangential vectors} as
\begin{equation}
    \Vec{X}_{ij} := \frac{1}{2}\bra{z_{i}}\Vec{\sigma}\ket{z_{j}}, \hspace{2.5mm} \Vec{S}_{ij} := \frac{1}{2}\boxbra{z_{i}}\Vec{\sigma}\ket{z_{j}}.
    \label{DEF generalized vectors}
\end{equation}
These vectors have been shown to appear when computing the evolution of the normal vectors to the faces of a polyhedron in the two-vertex model \cite{Cendal:2024uzu}.
Note that the diagonal part $\vec{X}_{i} = \vec{X}_{ii}$ is the flux vector normal to the $i$th face of the polyhedron and
 \begin{equation}
    \vec{S}_{i} := \vec{S}_{ii}=:\vec{G}_i+i\vec{F}_i,
    \label{DEF FyG}
 \end{equation}
with $\vec{G}_i,\vec{F}_i \in \mathbb{R}^3$, is tangent to the face. $\vec{X}_i, \vec{F}_i$ and $\vec{G}_i$ satisfy
\begin{equation}
    \vec{X}_i \cdot \vec{G}_{i} = \vec{X}_i \cdot \vec{F}_{i}= \vec{G}_i \cdot \vec{F}_{i}=0.
\end{equation}
Furthermore
\begin{equation}
|\vec{X}_{i}|=|\vec{F}_{i}|=|\vec{G}_{i}|=A_i,
\end{equation}
where, let us recall, $A_i$ is the area of the $i$th face.
Therefore, given the spinor $\ket{z_i}$ attached to the $i$th face of a polyhedron, we can
define the associated frame basis (see figure \ref{figure_frame_basis}) as the set $\{\Vec{X}_{i},\Vec{G}_{i},\Vec{F}_{i}\}$ with the orientation
\begin{equation}
    \Vec{X}_{i}\times\Vec{G}_{i} = A_{i}\Vec{F}_{i}.\label{productos vectoriales}
\end{equation}
The vectors $\vec{F}_{i}$ were originally introduced in \cite{Freidel:2013fia} to characterize the concept of framed polyhedra, used to describe the semiclassical behavior of coherent states on a graph \cite{Freidel:2009ck,Freidel:2010tt,Bonzom:2012bn}.

The frame basis at a polyhedron face contains all the information about the spinor attached to it, except its global sign. This is a different way to encode the same information provided by the twisted geometries variables, including the twist angle, since
all those quantities are independent of the global sign. This means that the frame bases at two contiguous faces of two contiguous polyhedra contain all the information about how one polyhedron should be rotated towards the other in order to \textit{glue} these faces. The following relations will be useful to make this information explicit.

Taking into account the definitions \eqref{DEF generalized vectors} and \eqref{DEF FyG},
the well known identities \cite{tisza2005applied}
\begin{align}
    \ket{z_i}\bra{z_i} &= |\Vec{X}_i| \mathbb{I} + \Vec{X}_i\cdot\Vec{\sigma},\\
  \boxket{z_i}\boxbra{z_i} &= |\Vec{X}_i|\mathbb{I} - \Vec{X}_i\cdot\Vec{\sigma},\\
    \ket{z_i}\boxbra{z_i} &= \Vec{G}_i\cdot\Vec{\sigma} + i\Vec{F}_i\cdot\Vec{\sigma},
  \end{align}
  are recovered, from where it follows, c.f. \eqref{DEF holonomia}
\begin{align}
    -g_i X^{s}_i g^{-1}_i &= X^{t}_i,\\ -g_i G^{s}_i g^{-1}_i &= G^{t}_i, \\ -g_i F^{s}_i g^{-1}_i &= -F^{t}_i.
\end{align}

\begin{figure}
    \centering
    \includegraphics[width=0.4\textwidth]{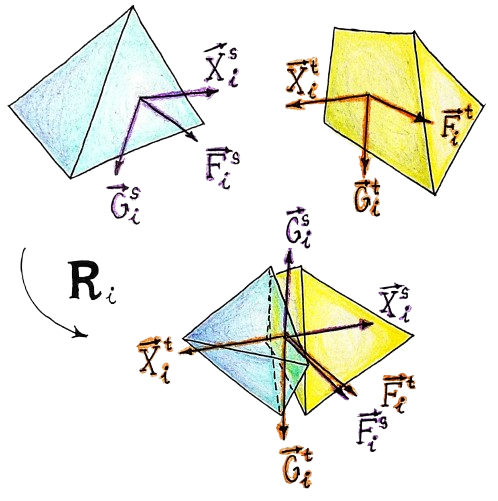}
    \centering
    \caption{Representation of the transformation \eqref{rotation_matrices}, that establishes the way two contiguous faces are glued together. The frame basis at the source node of a link $i$ is rotated until the vectors $\vec{X}^{s}_i, \vec{X}^{t}_i$ are parallel with opposite directions and the vectors $\vec{F}^{s}_i, \vec{F}^{t}_i$ are parallel with the same direction. This transformation can also be determined from the normal vectors $\vec{X}^{s}_i, \vec{X}^{t}_i$ if the twist angle $\xi$ is known (see e.g. \cite{freidelTwistedGeometriesGeometric2010a}).}
    \label{figure_gluing}
\end{figure}

The action of the holonomy at 
a link connecting two faces is thus a rotation $\mathbf{R}_i\in  \text{SO}(3)$ given by
\begin{equation}
    {R_i}^{IJ} = -\frac{1}{A_i^2}\left(X^{t I}_iX^{s J}_i + G^{t I}_iG^{ s J}_{i} - F^{ t I}_{i}F^{ s J}_{i}\right),
    \label{rotation_matrices}
\end{equation}
that transforms
the frame basis at the source face $\{\Vec{X}^{s}_i, \Vec{G}^{s}_i, \Vec{F}^{s}_i\}$ into $ \{-\Vec{X}^{t}_i,-\Vec{G}^{t}_i, \Vec{F}^{t}_i\}$ (see figure \ref{figure_gluing}). 
Therefore,  given the frame bases at the source and target nodes of all the links of the graph, all the polyhedra can be directly reconstructed from the normal vectors $\vec{X}_i$, and the way of gluing them is specified by the rotation matrices \eqref{rotation_matrices}. This information concerning the gluing is, precisely, what is encoded in the twist angle $\xi$ when using the twisted geometry variables. However, there is more than one valid definition of twist angles, and, depending on that, their precise  interpretation in terms of the rotation just described may vary \cite{freidelTwistedGeometriesGeometric2010a,freidelTwistorsTwistedGeometries2010}.

The $\U(1)$ transformation \eqref{U1} on the spinors at a link induces the following transformation on the two frame bases linking the two faces  given by 
\begin{equation}
    \mathbf{M}^{s} \to \mathbf{G}(\phi) \mathbf{M}^s, \hspace{2.5mm} \mathbf{M}^{t} \to \mathbf{G}(-\phi) \mathbf{M}^t,
\end{equation}
 with
 \begin{align}
      \mathbf{M} &:= \begin{pmatrix}
        X^{1} & X^{2} & X^{3} \\
        G^{1} & G^{2} & G^{3} \\
        F^{1} & F^{2} & F^{3}
      \end{pmatrix}\end{align}
   for each source $s$  and target $t$ vectors, and
    \begin{align}
    \mathbf{G}(\phi) &:= \begin{pmatrix}
        1 & 0 & 0 \\
        0 & \cos{\left(2\phi\right)} & -\sin{\left(2\phi\right)} \\
        0 & \sin{\left(2\phi\right)} & \cos{\left(2\phi\right)}
    \end{pmatrix}.
  \end{align}
  Observe that this transformation is a rotation around the axis
defined by the normal vectors $\vec{X}_i^{s}$ and $\vec{X}_i^{t}$ , but with opposite
direction for the source and target. The dependence of the gauge
rotation of the frame on $2\phi$ reflects the fact
that the phase of the spinors can be determined from the frame basis
up to the sign of the spinor.

\section{Two-vertex model}

\begin{figure}
    \centering
    \includegraphics[width=0.4\textwidth]{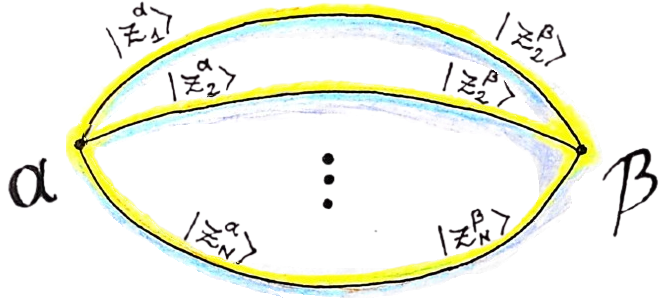}
    \centering
    \caption{The two-vertex graph is given by the nodes $\alpha, \beta$ and $N$ links connecting them. Attached to each node and link there is an associated spinor, labeled by the link it is attached to (lower index) and the node it belongs to (upper index).}
    \label{figure_two-vertex}
\end{figure}

The study of the so called two-vertex model,  given by a fixed graph with two vertices, $\alpha$ and $\beta$, and $N$ links (see figure \ref{figure_two-vertex}) has been very fruitful in order to implement dynamics and to explore cosmological scenarios within the full theory \cite{Rovelli_Vidotto_SteppingOutHomogeneity2008,Borja:2011ha,Borja_Freidel_Garay_Livine_2010_Return, Borja_Diaz-Polo_Garay_Livine_2010_Dynamics,Livine_Martin-Benito_ClassicalSettingEffective2013, Aranguren_Garay_Livine_2022, Cendal:2024uzu,Diego_Garays_2025}.

Using the twisted geometries framework, the model corresponds to two polyhedra with $N$ faces, where the links relate faces with same area (although possibly different shapes). Both the quantum mechanical and the classical dynamics of the model have been extensively studied in the literature, and reduced sectors with cosmological interpretation have been explored \cite{Borja_Diaz-Polo_Garay_Livine_2010_Dynamics,
Borja_Freidel_Garay_Livine_2010_Return,Livine_Martin-Benito_ClassicalSettingEffective2013,
Aranguren_Garay_Livine_2022,Cendal:2024uzu,Diego_Garays_2025}.

\subsection{Dynamics revisited}
The dynamics introduced in \cite{Borja_Diaz-Polo_Garay_Livine_2010_Dynamics,
Borja_Freidel_Garay_Livine_2010_Return}, and extensively used in \cite{Livine_Martin-Benito_ClassicalSettingEffective2013,
Aranguren_Garay_Livine_2022,Cendal:2024uzu}, is given by the Hamiltonian
\begin{equation}
    H = \lapseM \sum_{i,j = 1}^{N}\left[\lambda E_{ij}^{\alpha}E_{ij}^{\beta} + \Re(\gamma F_{ij}^{\alpha}F_{ij}^{\beta})\right],
    \label{Hamiltonian_equation1}
\end{equation}
where 
$\lapseM$ is the Lagrange multiplier, $\lambda \in \mathbb{R}$ and $\gamma \in \mathbb{C}$ are coupling constants, and the observables $E_{ij}$ and $F_{ij}$ are defined at each vertex as \cite{Livine2011} 
\begin{equation}
  E_{ij} := \braket{z_i}{z_j}, \quad F_{ij} := [z_i\ket{z_j}.\label{invariantes_su2}
\end{equation}
These observables are especially suitable in order to write the dynamics
because they are $\SU(2)$ invariant quantities.
Indeed, we can consider a
  generalisation of \eqref{Hamiltonian_equation1} by adding an arbitrary function
  $\af(A^{\alpha\beta})$, with
  \begin{align}
    A^{\alpha\beta}:=&\frac{1}{8}\sum_{i,j=1}^N \left[|E^\alpha_{ij}|^2+|F^\alpha_{ij}|^2+|E^\beta_{ij}|^2+|F^\beta_{ij}|^2\right]\nonumber\\
    =&\frac{1}{2}\sum_{i,j=1}^N \left[A^\alpha_i A^\alpha_j+A^\beta_i A^\beta_j\right],
  \end{align}
  which reduces to $A^{\alpha\beta}\equalc \sum_{ij} A^\alpha_iA^\alpha_j\equalc \sum_{ij} A^\beta_iA^\beta_j=:A^2$, i.e. the square
  of the total area of both polyhedra,
  when the matching constraint \eqref{DEF matching constraint} is imposed.
  Let us stress that this leads to the generalization of the Hamiltonian $\Hfull$ introduced in \cite{Cendal:2024uzu} after the $\U(N)$ reduction.
  To keep that notation, in this work we therefore consider the Hamiltonian
  \begin{equation}
    \Hfull=H+2\lapseM A^{\alpha\beta}\af(A^{\alpha\beta}).\label{Hamiltonian_equation2}
  \end{equation}

The Hamiltonian \eqref{Hamiltonian_equation2} is real, symmetric under the interchange of nodes, and Poisson commutes  with the closure \eqref{DEF closure constraint} and matching \eqref{DEF matching constraint} constraints, which makes it gauge invariant. Moreover, the Hamiltonian is a constraint (analogously to the Hamiltonian constraint of LQG and GR) and hence it is first class together with the matching and closure constraints. 
The time evolution of the model is then generated by this Hamiltonian via Poisson brackets.
In the remainder we will use $t$ to parametrize the time evolution, so that
  we will write
  \[
    \frac{df}{dt}:=\{f,\Hfull\}.
  \]
for any function $f$ in phase space.

Next
we give a full characterization of the evolution of polyhedra in the two-vertex model using the frame basis developed in section \ref{Sec_Frame_basis}, in order to complement and support the concrete cosmological results obtained in \cite{Cendal:2024uzu}.


\subsection{Evolution of the frame basis}

The time evolution of the normal vectors $\vec{X}^{\alpha,\beta}_i$ under $H$ was
studied in \cite{Cendal:2024uzu}. Now, since $\{A^{\alpha\beta},\vec X_i^\nu\}=0$
for $\nu=\alpha,\beta$, we have
\begin{align}
  \frac{d \vec{X}^\alpha_{i}}{dt}=&\{\vec{X}^\alpha_{i},\Hfull \} \nonumber\\
  = &2\lapseM\Im\sum_{j}\left(\lambda E^{\beta}_{ij} \Vec{X}^\alpha_{ij} + \gamma F^{\beta}_{ij}\Vec{S}^\alpha_{ij}\right)=:\vec{\Delta}_{i}^\alpha ,
      \label{evolutionVectores}
\end{align}
and analogously for the node $\beta$.
Similarly, a direct computation leads first to
  \begin{align*}
  \{A^{\alpha\beta},\vec S^\nu_i\}=iA^\nu \vec S_i^\nu,
  \end{align*}
from where it is not difficult to show that
the tangent vectors $\vec{F}_i$ and $\vec{G}_i$ evolve under $\Hfull$ according to
\begin{align}
  \frac{d\Vec{G}^\alpha_{i}}{dt} = & 
  2\lapseM\Im\sum_{j}\left(\lambda E_{ij}^{\beta}\Vec{S}^\alpha_{ij}-{\gamma}{F}_{ij}^{\beta}\Vec{X}^\alpha_{ij}\right) \nonumber \\
  &+2\lapseM \left(\af(A^{\alpha\beta})+ A^{\alpha\beta}\af'(A^{\alpha\beta})\right) A^\alpha \vec F_i^\alpha,
  \label{evolutionH} \\
  \frac{d\Vec{F}^\alpha_{i}}{dt}  = &-2\lapseM\Re\sum_{j}\left(\lambda E_{ij}^{\beta}\Vec{S}^\alpha_{ij}+{\gamma}{F}_{ij}^{\beta}\Vec{X}^\alpha_{ij}\right)\nonumber\\
  &-2\lapseM \left(\af(A^{\alpha\beta})+ A^{\alpha\beta}\af'(A^{\alpha\beta})\right)A^\alpha \vec G_i^\alpha,
    \label{evolutionF}
\end{align}
where the prime denotes derivative with respect to the argument,
  and analogously for the node $\beta$. Observe that $A^\nu:=\sum_iA_i^\nu$
  is the area of the polyhedron corresponding to the node $\nu$, and $A^\alpha\equalc A^\beta$ is thus the total area $A$
common to the two polyhedra.

On the other hand, a straightforward calculation shows that the generalized normal and tangent vectors can be written with respect to the frame basis as
\begin{align}
  \Vec{X}_{ij} &= \frac{1}{2A_{i}}\left({E}_{ij}\Vec{X}_{i} + F_{ij}\Vec{G}_{i} -iF_{ij}\Vec{F}_{i}\right),\\
  \Vec{S}_{ij} &= \frac{1}{2A_{i}}\left(- F_{ij}\Vec{X}_{i} + {E}_{ij}\Vec{G}_{i} + i{E}_{ij}\Vec{F}_{i}\right),
    \label{C1}
\end{align}
on each node.
Therefore, in terms of  the set of real quantities $\{a_i,b_i,c_i,d_i,e_i,f_i,g_i,h_i\}$ defined by
\begin{align}
        &\sum_{j} E_{ij}^{\alpha}E_{ij}^{\beta} =: a_{i} + ib_{i},
        &\sum_{j} F_{ij}^{\alpha}E_{ij}^{\beta} =: c_{i} + id_{i},\label{abecedario1}\\
        &\sum_{j}F_{ij}^{\alpha}F_{ij}^{\beta} =: e_{i} + if_{i},
        &\sum_{j}E_{ij}^{\alpha}F_{ij}^{\beta} =: g_{i}+ih_{i},\label{abecedario2}
      \end{align}
    and using $\gamma_R$ and $\gamma_I$ for the real and imaginary parts of $\gamma$,
    we can express the evolution of the frame bases of the node $\alpha$  as
    \begin{equation}
        \frac{d}{dt}\begin{pmatrix}
          \Vec{X}^\alpha_{i} \\
          \Vec{G}^\alpha_{i} \\
          \Vec{F}^\alpha_{i}
        \end{pmatrix} = \mathfrak{M}^\alpha_{i} \begin{pmatrix}
          \Vec{X}^\alpha_{i} \\
          \Vec{G}^\alpha_{i} \\
          \Vec{F}^\alpha_{i}
        \end{pmatrix}
        \label{matricial}
      \end{equation}
    with
    \begin{widetext}
     \begin{align}
              &\mathfrak{M}^\alpha_{i} =
    \frac{\lapseM}{A_{i}}
    \begin{pmatrix}
      \lambda b_i  -\gamma_R f_i - \gamma_I e_i
      &\lambda d_i+\gamma_R h_i  +\gamma_I g_i
      & -\lambda c_i +\gamma_R g_i-\gamma_I h_i \\
      -(\lambda d_i+\gamma_R h_i  +\gamma_I g_i)
      & \lambda b_i  -\gamma_R f_i - \gamma_I e_i
      & \lambda a_i + \gamma_R e_i- \gamma_If_i +2(\af + A^2\af') A_i A \\
      \lambda c_i -\gamma_R g_i+\gamma_I h_i
      & -(\lambda a_i + \gamma_R e_i- \gamma_If_i +2(\af + A^2\af') A_i A)
      &   \lambda b_i  -\gamma_R f_i - \gamma_I e_i
  \end{pmatrix}.\label{Mi}
 \end{align}
For the node $\beta$, the analogous computation leads to
  \eqref{matricial} changing $\alpha$ by $\beta$ and with
\begin{align}
              &\mathfrak{M}^\beta_{i} =
    \frac{\lapseM}{A_{i}}
  \begin{pmatrix}
      \lambda b_i  -\gamma_R f_i - \gamma_I e_i
      &\lambda h_i+\gamma_R d_i  +\gamma_I c_i
      & -\lambda g_i +\gamma_R c_i-\gamma_I d_i \\
      -(\lambda h_i+\gamma_R d_i  +\gamma_I c_i)
      & \lambda b_i  -\gamma_R f_i - \gamma_I e_i
      & \lambda a_i + \gamma_R e_i- \gamma_If_i +2(\af + A^2\af') A_i A \\
      \lambda g_i -\gamma_R c_i+\gamma_I d_i
      & -(\lambda a_i + \gamma_R e_i- \gamma_If_i +2(\af + A^2\af') A_i A)
      &   \lambda b_i  -\gamma_R f_i - \gamma_I e_i
    \end{pmatrix}.\label{Mi_beta}
        \end{align}
\end{widetext}
Notice that these matrices (for each node and for each link $i$) can be decomposed as the sum of a term proportional to the identity, describing the expansion of the frame basis, and an antisymmetric term, describing its rotation.
No stress appears since the vectors constituting the frame basis are orthogonal and have the same modulus by construction.

Let us stress that this result describes the evolution in time of the twisted geometries
information in the general two-vertex model.
In particular, note also that the evolution of the normal vectors is unaffected by
  the term with $\af(A^{\alpha\beta})$ in the Hamiltonian.

\subsection{Evolution of the planar angles}
The faces of the convex polyhedra described by the twisted geometries are polygons whose shapes are determined by the planar angles between their edges. 
These angles can thus be used to characterize the shape of the polyhedra and analyze the evolution.
Let us consider the unit vectors
\begin{equation}
    \Vec{u}_{ij} =  \frac{\left(\Vec{X}_{i}\times\Vec{X}_{j}\right)}{|\Vec{X}_{i}\times\Vec{X}_{j}|}.
    \label{uij}
\end{equation}

\begin{figure}
    \centering
    \includegraphics[width=0.4\textwidth]{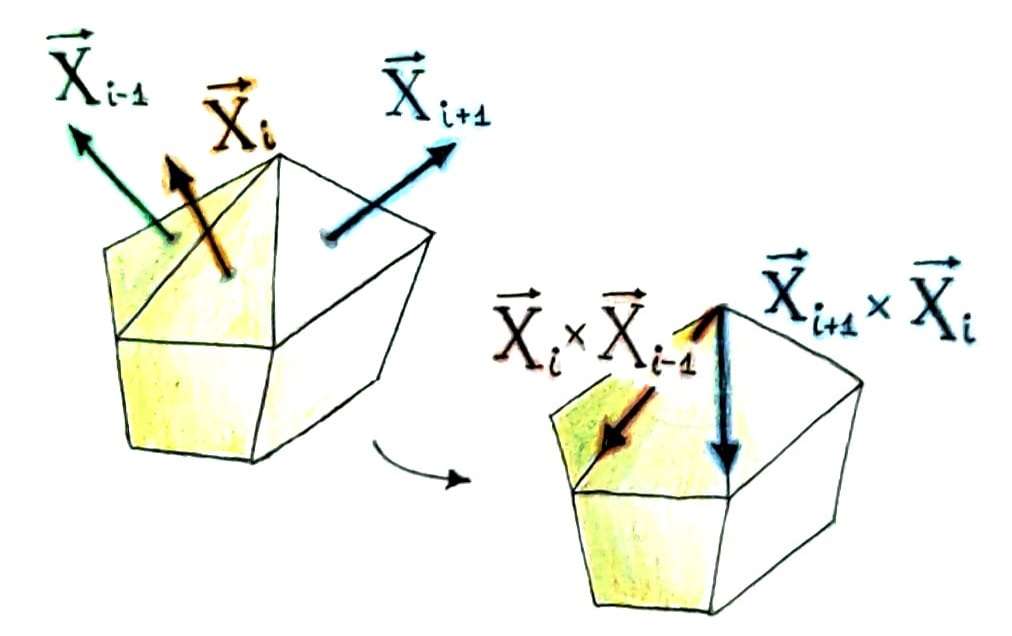}
    \centering
    \caption{Representation of a polyhedron, the normal vector to one of its faces, $\Vec{X}_{i}$, and the normal vectors of two adjacent faces, $\Vec{X}_{i-1}$ and $\Vec{X}_{i+1}$. The vector products $\Vec{X}_{i}\times\Vec{X}_{i-1}$ and $\Vec{X}_{i+1}\times\Vec{X}_{i}$ have the same direction as the edges between the faces. The angle between these vectors provides the internal angle of the face of the polyhedron.}
    \label{figure_planar_angles}
\end{figure}

If the normal vectors $\vec{X}_i$ and $\vec{X}_j$ are associated to
contiguous faces of the polyhedron, the vector $\vec{u}_{ij}$ has the
direction of the shared edge between the faces.
Therefore, we can take the vectors along two edges of a polygon meeting at one of its vertices to compute the cosine of the planar angle between them. 

Let us then denote
by $\Vec{X}_{i-1}$ and $\Vec{X}_{i+1}$
the normal vectors to two faces which are adjacent to the face $i$ and share a common vertex (see figure \ref{figure_planar_angles}).
The cosine $\varepsilon_i$ of the planar angle between the two edges is thus given by
\begin{equation}
    \varepsilon_{i} = \Vec{u}_{i(i-1)}\cdot\Vec{u}_{(i+1)i}.
\end{equation}
The time evolution of these quantities can now be computed directly from the evolution equations
of the frame basis. That yields
\begin{align}
  \frac{d\varepsilon_{i}}{dt}= 
  \left(\Vec{u}_{i(i-1)}\cdot\Vec{v}_{(i+1)i} + \Vec{u}_{(i+1)i}\cdot\Vec{v}_{i(i-1)}\right)\nonumber\\
  - \varepsilon_{i} \left(\Vec{u}_{i(i-1)}\cdot\Vec{v}_{i(i-1)} + \Vec{u}_{(i+1)i}\cdot\Vec{v}_{(i+1)i}\right),
    \label{epsilonCORCHETE}
\end{align}
where
\begin{align}
    \Vec{v}_{ij} := \frac{\Vec{\Delta}_{i}\times\Vec{X}_{j} - \Vec{\Delta}_{j}\times\Vec{X}_{i}}{|\Vec{X}_{i}\times\Vec{X}_{j}|},
    \label{vij}
\end{align}
and $\Vec{\Delta}_{i}$ is defined in \eqref{evolutionVectores}.

As expected from the numerical study of the evolution of the two-vertex model carried out in \cite{Aranguren_Garay_Livine_2022} the planar angles between the edges of a polyhedron will evolve in general in a non-trivial manner,
thus modifying its shape in many different possible ways.
In the next section we show that in the $\U(N)$-reduced sector that evolution turns out
to be trivial.

\section{$\U(N)$ symmetry reduced sector}
\label{sec:unsector}
The cosmological results found in \cite{Livine_Martin-Benito_ClassicalSettingEffective2013,Cendal:2024uzu} rely on 
considering a reduced sector of the model, defined by
imposing a global $\U(N)$ symmetry on the graph \cite{Borja_Diaz-Polo_Garay_Livine_2010_Dynamics,
Borja_Freidel_Garay_Livine_2010_Return}. The $\U(N)$ transformations are
generated via Poisson brackets by the quantities
\begin{equation*}
  \mathcal{E}_{ij} = E_{ij}^{\alpha} - E_{ji}^{\beta},
\end{equation*}
and it can be checked that these have vanishing Poisson
brackets with the Hamiltonian and closure constraints.
 On the other hand, the Poisson bracket of $\mathcal{E}_{ij} $
with the matching constraint $\mathcal{C}$ yields an expression proportional to
  $\mathcal{E}_{ij}$ (see e.g. \cite{Diego_Garays_2025}).
  As a result, the imposition of $\mathcal{E}_{ij}= 0$
  ensures that the matching constraint is satisfied along evolution.
This imposition defines the $\U(N)$-reduced sector of the model.

The $\U(N)$ reduction of the two-vertex model imposes the relation
on the spinors at the source and target nodes given by
\begin{equation}
    \ket{z^\beta_i} = e^{-i\varphi/2}\ket{\bar{z}^\alpha_i}\label{U(N) condition}
\end{equation}
with $\varphi\in[-\pi,\pi]$ an arbitrary phase, equal for all links. 
The Hamiltonian \eqref{Hamiltonian_equation2}, with \eqref{Hamiltonian_equation1},
reduces then to the expression
\begin{equation}
    \Hfullred = 2\lapseM A^2\left(\lambda +\Re(\gamma e^{-i\varphi})\right)+2\lapseM A^2 \af(A^2),
    \label{omega}
\end{equation}
where, let us recall, $A$ is the total area of any of the two polyhedra of the model.\footnote{For definiteness, let us note that $\af(A^2)$ relates with the function $g(A)$ used in
  \cite{Cendal:2024uzu} by $\af(A^2)=g(A)$.}

As shown in \cite{Cendal:2024uzu}, the phase $\varphi$ equals the twist angle $\xi_i$
for all links $i$ in this $\U(N)$-reduced sector, and therefore the twist is link-independent.
The twist $\xi=\varphi$ and the total area $A$ turn out to be conjugate variables
  with $\{\varphi,A\}=1$
  \cite{Borja_Freidel_Garay_Livine_2010_Return,Cendal:2024uzu}.
  The equations of motion (on shell) read
  \begin{align}
    &\frac{d\varphi}{dt}=\{\varphi,\Hfullred\}=4\lapseM A^3 \af'(A^2),\label{phi dot}\\
    &\frac{dA}{dt}=\{A,\Hfullred\}= -2\lapseM A^2\Im(\gamma e^{-i\varphi}).\label{A dot pre}
  \end{align}
  The vanishing of the Hamiltonian constraint,
  $\Hfullred=0$, fixes one of the constants
  of integration. The other constant of integration can be taken to correspond to
  a shift in $\varphi$ that can be fixed by demanding a real-valued coupling constant
  $\gamma$. As a result we can assume without loss of generality that $\gamma\in\mathbb{R}$,
  and the system of equations we will take in what follows is
  \begin{align}
    \Hfullred=\quad&2\lapseM A^2\left(\lambda +\gamma \cos\varphi+\af(A^2)\right)=0,\label{vanish constraint}\\
              &\frac{dA}{dt}= 2\lapseM A^2\gamma \sin\varphi.\label{A dot}
   \end{align}
  
One of the main conclusions in \cite{Cendal:2024uzu} is based on the computation of the evolution of the volume of the two polyhedra.
Since there is no known closed analytic form of the volume of a polyhedron of more than four faces in terms of their normal vectors,
an approximate volume must be used for the general case. A convenient candidate is defined in terms of the geometric quadrupole of the polyhedron \cite{goellerProbingShapeQuantum2018}
\[
  T^{IJ}=\sum_i \frac{1}{|\vec X_i|}X_i^I X_i^J,
\]
from where the approximate volume $\tilde V$ is defined by
\[
  \tilde V=\sqrt{\frac{4\pi}{3}\det(T)}.
\]
In \cite{Cendal:2024uzu} it was shown, in particular, that $\tilde V$ satisfies the equation
\begin{equation}\label{expansion}
  \frac{1}{\tilde{V}} \frac{d\tilde V}{dt}=\frac{1}{\tilde{V}} \{\tilde V, \Hfullred \}=
  3\lapseM A\gamma \sin \varphi=:\expansio,
\end{equation}
so that it holds, c.f. \eqref{A dot}
\begin{equation}
  \frac{1}{\tilde{V}} \frac{d\tilde V}{dt}=\frac{3}{2}\frac{1}{A} \frac{dA}{dt} \implies \tilde V=\tilde{a}_0A^{3/2}
  \label{Vtilde(t)}
\end{equation}
for some constant $\tilde{a}_0$.

This result suggested that polyhedra could be undergoing
a uniform rescaling of its dimensions, i.e. an homothetic expansion,
and an indirect evidence was given by the fact that the dynamics resembles,
in this case, that of the FLRW models  \cite{Cendal:2024uzu}.
However, a direct proof was still missing.
Moreover, although $\tilde V$ is proportional to the real volume $V$ for polyhedra of four faces
  \cite{goellerProbingShapeQuantum2018}, the validity of the approximation $\tilde V$
  still required further investigation.

  In the following subsection we are going to show that in the $\U(N)$-reduced sector of the two-vertex model
  polyhedra indeed experiment an homothetic expansion,
  and that, as a consequence, the approximate
  volume $\tilde V$ is proportional to the volume $V$ of the polyhedron of an arbitrary number of faces.

\subsection{Homothetic expansion of the polyhedra}

In order to prove that, we take the frame basis framework introduced in section \ref{Sec_Frame_basis}. In particular, imposing \eqref{U(N) condition} on \eqref{abecedario1}-\eqref{abecedario2} yields
\begin{align}
  a_i &= 2 A_{i}A ,\\
  e_i &= 2A_{i}A \cos\varphi,\\
  f_i &= - 2A_{i}A \sin\varphi,   
\end{align}
and $b_i = c_i = d_i = g_i = h_i = 0$. Therefore, applying the
  vanishing of the Hamiltonian constraint \eqref{vanish constraint},
  the evolution matrices for the frame basis relative to each link $i$ and nodes $\alpha$ and $\beta$,
  \eqref{Mi}-\eqref{Mi_beta}, reduce to
\begin{equation}
    \mathfrak{M}^\alpha_i =\mathfrak{M}^\beta_i=  \begin{pmatrix}
        \frac{2}{3}\expansio & 0 & 0 \\
        0 & \frac{2}{3}\expansio &  2\lapseM A^3\af '  \\
        0 &  -2\lapseM A^3\af ' & \frac{2}{3}\expansio
    \end{pmatrix}.
  \end{equation}

Two important comments are in order.
First, all the matrices $\mathfrak{M}^\alpha_i $ and $\mathfrak{M}^\beta_i $ are equal,
they do not depend neither on the link (face) $i$ nor on the node.
  Therefore, the evolution of the frame basis is independent of the face and of the polyhedron.
  Second, the antisymmetric part of the matrix, that produces a rotation of the frame
  with axis along the normal vector $\vec{X}_i$ on each face,
  equals  $\dot\varphi /2$, which is proportional to $\af'$ by \eqref{phi dot}.
  To sum up, the evolution of all the frame bases is given by
\begin{equation}
  \frac{d}{dt}\begin{pmatrix}
    \Vec{X}_{i} \\
    \Vec{G}_{i} \\
    \Vec{F}_{i}
  \end{pmatrix} = \frac{2}{3}\expansio \begin{pmatrix}
    \Vec{X}_{i} \\
    \Vec{G}_{i} \\
    \Vec{F}_{i}
  \end{pmatrix}
  +\begin{pmatrix}
      0 & 0 & 0 \\
      0 & 0 &\dot \varphi/2 \\
      0 & -\dot \varphi/2 &0 
    \end{pmatrix}\begin{pmatrix}
      \Vec{X}_{i} \\
      \Vec{G}_{i} \\
      \Vec{F}_{i}
    \end{pmatrix}.
  \label{frame evolution}
\end{equation}

As a consequence, in the $\U(N)$-reduced version of the two-vertex model
the vectors $\Vec{X}_{i}$, $\Vec{G}_i$ and $\Vec{F}_{i}$ evolve in time
both varying their module according to the expansion $\expansio$,
and rotating according to $\dot \varphi\propto \af'$.
Notice that since the expansion and rotation are independent of the face,
all the faces modify their area at the same rate.

Moreover, introducing \eqref{frame evolution} in equation \eqref{vij}
yields a very simple relation between the vectors $\Vec{u}_{ij}$ and $\Vec{v}_{ij}$, given by
\begin{equation}
    \Vec{v}_{ij} = \frac{4}{3}\expansio\Vec{u}_{ij}.
\end{equation}
This can be, in turn, introduced into \eqref{epsilonCORCHETE}
to obtain 
\begin{equation}
   \frac{d\varepsilon_{i}}{dt}:=  \left\{\varepsilon_{i}, H\right\} = 0.
     \label{epsilon0}
   \end{equation}
This shows, as already suspected in \cite{Cendal:2024uzu},   that \emph{polyhedra undergo a uniform rescaling of their
dimensions without changing their shape}. Indeed, equations \eqref{epsilon0}
are
satisfied for any of the planar angles of each face of the polyhedron we are taking.
That is,  \eqref{epsilon0} corresponds to a homothetic evolution of the polyhedra.

In turn, the homothetic expasion of the polyhedra implies that their volume $V$ and area $A$
  must be related through
  \begin{equation}
    V\propto A^{3/2}.\label{V propto A}
  \end{equation}
  By comparing this relation with  \eqref{Vtilde(t)} we can ensure that, at least in the
reduced $ \U(N)$ sector of the model, it holds
\[
  \tilde V\propto V
\]
with a constant proportionality factor.
This means, in particular,
that \emph{the approximate volume $\tilde{V}$ in terms of the quadrupole moment has the same
  behavior as the real volume $V$ of the polyhedra.}
Explicitly, the combination of \eqref{V propto A} with \eqref{Vtilde(t)} allows us to write
\begin{equation}
  V=a_0A^{3/2} \label{V evolution}
\end{equation}
for some constant $a_0$ that only depends on the initial conditions. Finally, observe that the expansion $\expansio$ associated to the volume $\tilde{V}$
corresponds to the expansion of the real volume $V$, that is, c.f. \eqref{expansion},
\begin{align}\label{expansion loop}
  \frac{1}{V} \frac{d V}{dt}=\expansio=3\lapseM A \gamma \sin\varphi.
\end{align}

\section{Cosmological interpretation: emergence of FLRW}\label{Sec_cosmo}

The cosmological interpretation of the $\U(N)$-reduced sector of the two-vertex model
dates back to the work in \cite{Borja_Diaz-Polo_Garay_Livine_2010_Dynamics,
Borja_Freidel_Garay_Livine_2010_Return}, and it was further explored in the works
presented in \cite{Livine_Martin-Benito_ClassicalSettingEffective2013,
Aranguren_Garay_Livine_2022} and, more recently, in \cite{Cendal:2024uzu}.
In particular, the result obtained in \cite{Cendal:2024uzu} is based on two points.

The first point is the recovery of the LQC equations
for a perfect fluid compatible with the RW geometry, with scale factor
$a(t)$ proportional to $\sqrt{A(t)}$, 
that reduces indeed to the Friedmann and continuity equations
for a perfect fluid compatible with FLRW.

For preciseness and completeness let us review the arguments
given in \cite{Cendal:2024uzu}.
Under the change of variables (it is a canonical transformation)
\begin{equation}\label{can_transf}
  a=\sqrt{\beta}\sqrt{A},\quad \pi_a=-\frac{2}{\sqrt{\beta}}\sqrt{A}\varphi,
\end{equation}
for an arbitrary fixed constant $\beta>0$ (that provides the Barbero-Immirzi ambiguity),
plus the convenient redefinitions
\[
  f(a):=2\frac{1}{\beta^2 a^2}\af(A^2(a)),\quad \lapseM=:\frac{1}{a^3}\lapseN,
\]
the system of equations \eqref{vanish constraint}-\eqref{A dot} can be written as
\begin{align}
  &\frac{1}{\lapseN}\frac{da}{dt} 
    =-\frac{\gamma}{\beta} \sin\left(\frac{\beta\pi_a}{2a}\right),\label{dot a}\\
  &\frac{1}{a^2}\left(\frac{1}{\lapseN} \frac{da}{dt}\right)^2 
    =\left(\gamma f-\frac{\kappa}{a^2}\right)
    \left[1-\frac{\beta^2 a^2}{4\gamma}\left(f-\frac{\kappa}{\gamma a^2}\right)\right],\label{eq:Friedmann1}
\end{align}
where
\[
  \kappa= - \frac{2}{\beta^2}\gamma(\lambda+\gamma).
\]
Observe that the evolution operator reads $\lapseN^{-1} d/dt$,
  now taking into account the lapse function $\lapseN$,
  thus providing a normalized time evolution operator associated with the Hamiltonian of the two-vertex model.
  
The first equation of the system, \eqref{dot a},
determines $\pi_a$ algebraically,
while the second, \eqref{eq:Friedmann1}, is the equation for $a(t)$
once $f(a)$ (or $\af$) is given, which provides the link to LQC as explained
in \cite{Cendal:2024uzu}.

Consistently, in the small twist regime $|\varphi|\ll 1 $, which is equivalent to
$|\pi_a \beta/a|\ll1$,
the system reduces to
\begin{align}
  & \frac{1}{\lapseN}\frac{da}{dt} 
    =-\gamma\frac{\pi_a}{2a},\label{pi_a}\\
  & \frac{1}{a^2}\left(\frac{1}{\lapseN} \frac{da}{dt}\right)^2=\gamma f-\frac{\kappa}{a^2},\label{eq:friedmann_pre}
\end{align}
which is independent of $\beta$, as expected. The key fact is that the second
equation reduces this time to a form equivalent to the Friedmann equation
(with cosmological constant $\Lambda$)\footnote{Observe that terms
  in $\af$ proportional to $A$, which represent
  terms proportional to $A^3$ in the constraint $\Hfullred/\lapseM$
  translate to constant terms in $f$,
  and thus contribute to the Cosmological Constant,
  see \cite{Livine_Martin-Benito_ClassicalSettingEffective2013}.}
\[
  \frac{1}{a^2}\left(\frac{1}{\lapseN} \frac{da}{dt}\right)^2
  =\frac{8\pi G}{3}\varrho+\frac{\Lambda}{3} -\frac{\kappa}{a^2},
\]
for a FLRW model with scale factor $a$, spatial curvature $\kappa$
and perfect fluid energy density
\begin{equation}
  \varrho=\frac{3\gamma}{8\pi G} \left( f -\frac{\Lambda}{3\gamma}\right), \label{first_rho}
\end{equation}
where the gravitational constant is given by $G=3\gamma \mathcal{V}_\Sigma/8\pi$, with $\mathcal{V}_\Sigma$ a fiducial spatial volume needed in order to compare the dynamical equation of the two-vertex model with the Friedmann equation\footnote{
The Hamiltonian constraint used for the two-vertex model would come from an integration of the LQG Hamiltonian density over a fiducial spatial volume \cite{Cendal:2024uzu}.} \cite{Cendal:2024uzu}.

The pressure of the perfect fluid can be directly obtained from \eqref{first_rho}
assuming the continuity equation (in the RW geometry)
holds, i.e.
$
a \dot \varrho = -3\dot a (\varrho+p),
$
where the dot denotes a time derivative, to obtain
\[
  p = \frac{\gamma}{8\pi G}\left(\frac{\Lambda}{\gamma}-\frac{1}{a^2}\frac{d}{da} (a^3 f(a))\right).
\]
To sum up, this first point establishes
that this scenario 
is compatible with the FLRW model, that is,
a RW spacetime $(\mathbb{R}\times \Sigma_\kappa,g)$ with
\[
  g=-\lapseN ^2(t) \ed t^2 + a(t)^2 \ed \sigma_\kappa^2,
\]
where $\ed \sigma_\kappa^2$ denotes a metric of constant curvature $\kappa$,
with energy density $\varrho$ and pressure $p$ given by the
Einstein field equations (EFEs).

Let us note that the first point just reviewed establishes a ``physical''
relation between the $\U(N)$-reduced sector of the two-vertex model
and the FLRW model in the sense of the emergence of the Friedmann equation.

The second point, that we further develop next, is based on a set of
notions and  facts that one may use to argue a purely geometrical
correspondence 
with the RW geometry, that is, independently of the EFEs.
The notions of homogeneity and isotropy (of space)
in the  $\U(N)$-reduced sector of the two-vertex model
correspond to 
two respective facts. The notion of homogeneity
lies on the equality of the products $\vec X^\nu_i\cdot \vec X^\nu_j$
in both $\nu=\alpha,\beta$ (and thus the polyhedrons dual to the
two vertices $\alpha$ and $\beta$ are identical).
The notion of isotropy rests on the fact
that the only relevant degree of freedom is the total area $A$
(not the individual areas) and its conjugate variable $\varphi$
is the same on all links.
In short, 
both variables have  completely lost the information on the specific node or link \cite{Borja_Diaz-Polo_Garay_Livine_2010_Dynamics,
Borja_Freidel_Garay_Livine_2010_Return}.

These notions of homogeneity and isotropy of the
$\U(N)$-reduced sector of the two-vertex model
were further explored by the result found in \cite{Cendal:2024uzu} 
establishing that the evolution \eqref{A dot} of the total area $A(t)$ of the polyhedra  
compared to the evolution \eqref{expansion} of the approximate volume $\tilde V(t)$
is compatible  with an
homothetic expansion of the polyhedra (described in  \cite{Cendal:2024uzu}
as a ``homogeneous and isotropic evolution'').
The results of the previous section elevate that compatibility to
an assertion.
That is,  
the polyhedra indeed undergo an homothetic evolution.
As a direct consequence, the approximate volume $\tilde V(t)$ is proportional
  to the exact volume $V(t)$ (for arbitrary $N$), and therefore one can use
  $\tilde V(t)$
  to compute the exact expansion in this setting.

In the remainder of this section we elaborate a bit further on the purely
geometrical relation
that the $\U(N )$-reduced sector of the two-vertex model may have with the RW geometry.

Let us first point out that the consequence of the above
notions of homogeneity and isotropy, even taken to
the continuum limit, must be taken with care,
since homogeneity and isotropy of space
is not enough to characterise a RW geometry. To be precise,
consider a spacetime of the form $M=I\times \Sigma$ ($I$ and interval in $\mathbb{R}$)
foliated by the leaves $\Sigma_t=\{t\}\times \Sigma$, where $t$ is the projection onto $I$,
as it is common in the Hamiltonian formulation of GR.
Then, the condition that the leaves $\Sigma_t$ are homogeneous and isotropic,
i.e. of constant curvature $k(t)$ each,
is not enough to ensure that $M$ is RW (see e.g. \cite{Avalos2022}).
The idea is that homogeneity and isotropy of space is a property
of the intrinsic geometry of the leaves, and that to characterise RW
one further needs the information about how those leaves are embedded
in $M$, that is, we need information about the extrinsic curvature
of all $\Sigma_t$.

Fortunately, we argue next that the notion of isotropy in terms of the twist angle
provides the necessary information to relate
the $\U(N )$-reduced sector of the two-vertex model with the RW geometry.

The ingredient we need is the relation between the twist angle
and the extrinsic curvature of each leave $\Sigma_t$
in the twisted geometry formalism of loop gravity as described, e.g., in
\cite{freidelTwistedGeometriesGeometric2010a,rovelliGeometryLoopQuantum2010}.
Let $\gammam(\cdot,\cdot )$ be the metric of the discrete geometry on $\Sigma_t$
(let us ignore the dependence on $t$ of the metric not to overwhelm the notation).
The extrinsic curvature of the discrete  geometry
is constructed from the symmetric tensors  $K_{(\nu,i)}(\cdot,\cdot)$
on each face $i$ of the polyhedron $\nu$,
that satisfy the relations (up to a gauge in the twists) \cite{rovelliGeometryLoopQuantum2010}\footnote{
The introduction of the Barbero-Immirzi parameter $\beta$ in
  \cite{rovelliGeometryLoopQuantum2010} has been taken into account here in the canonical transformation \eqref{can_transf}. Therefore the parameter $\beta$ is already included in
$\xi$ here.}
$$
K_{(\nu,i)}(\cdot,N_{\nu,i})=\xi_i(\nu) \gammam|_{(\nu,i)}(\cdot, N_{\nu,i}),
$$
where $N_{\nu,i}$ denotes the (any)
vector normal to the face $i$, where $\gammam$ is evaluated. 
Now, if the twist does not depend on the link, 
then
$
  K_{(\nu,i)} =\xi(\nu) \gammam|_{(\nu,i)}.
  $
 Then, the pass to the continuum would allow us to write
 \begin{equation}\label{umbilic}
  K =\frac{1}{\ell}\xi  \gammam,
\end{equation}
where $\ell$ is a small length scale coming from the projection of  the continuous extrinsic curvature along the corresponding edge of the graph \cite{Freidel:2013fia,Cendal:2024uzu}.

In short, the above rough argument allows us to state that  the notion of isotropy coming from the equality of the twists
of all the links 
translates to the notion of umbilicity (first and second fundamental forms are proportional) of the leaves $\Sigma_t$ as embedded in $M$.

Umbilicity is a key point in the characterisation of RW that we can use now.
Indeed, RW can be (locally) characterised as the spacetime
$M=I\times \Sigma$ foliated by spacelike hypersurfaces $\Sigma_{\trw}$
with induced metric $\gammam_{\trw}$ and second fundamental form $K_{\trw}$ such that\footnote{This characterisation can be generalised in point (iii). As shown in Theorem 4.2 in \cite{RW_2024}, it suffices to ask that
  the gradient and Laplacian (with respect to $\gammam_{\trw}$, at each $\trw$) of $\expansionrw$ vanish along a single curve
transverse to the foliation.}
(see \cite{Krasinski,RW_2024})
\begin{itemize}
\item[(i)] each $(\Sigma_{\trw},\gammam_{\trw})$ is of constant curvature $k(\trw)$,
\item[(ii)] $\Sigma_{\trw}$ are umbilic, $K_{\trw}=\frac{1}{3}\expansionrw \gammam_{\trw}$ for some function $\expansionrw$,
\item[(iii)] $\expansionrw$ is constant on each leave,
\item[(iv)] the unit normal vector $U$ of the foliation $\{\Sigma_{\trw}\}$ is geodesic.
\end{itemize}
In terms of the flow $U$, points (ii) and (iii) above are equivalent to
demanding that $U$ is shear-free and its expansion, which is $\expansionrw$, only depends on $\trw$.
Observe that,  by contruction, the covector $\bmath{U}$ is proportional to $\ed \trw$.
The last point (iv) (once the rest hold) is then equivalent to demanding that the lapse function $\lapseRW$,
defined by $\bmath{U}=-\lapseRW \ed \trw$,
depends only on $\trw$ (see \cite{RW_2024}).

Therefore, given all the above, in the $\U(N)$-reduced sector of the two-vertex model
we can relate (i) to the homogeneous and isotropic notions on each leave,
(ii) and (iii) to the fact that all the twists are the same and equal $\varphi$,
that is, are constant on each leave, 
and (iv) to the fact that the lapse function is a function of $t$ only.
Let us finally stress that points (ii) and (iii) directly relate to the
homothetic evolution result found above.
In fact, the existence of an homotecy between leaves
  appears in other characterizations of RW  (see \cite{Avalos2022}).

For completeness, we end this section with a couple of consistency checks.
In order to proceed, let us stress first that
coordinates $\{\trw,x\}$ in RW can be chosen so that $U=\lapseRW^{-1}\partial_{\trw}$.
Note, however, that the derivatives along the flow $U$ have an intrinsic meaning, independent
on the coordinates chosen.
For that reason,
the evolution operator $\lapseN^{-1}d/dt$
in the two-vertex model, that corresponds to the unit normal derivative to the foliation,
will translate 
to $U(f)$ for functions
and, since $\lapseN$ is constant on each leave,
$[U,X]$ for vectors $X$ tangent to the foliation.

For the first consistency check we use that
in the $\U(N)$-reduced two-vertex model, equation \eqref{umbilic} with $\xi=\varphi$
and point (ii) establish
that
\begin{equation}\label{varphi_expansion}
\frac{1}{\ell}  \varphi=\frac{1}{3}\Theta.
\end{equation}
For the small twist regime, we use now \eqref{can_transf} and \eqref{pi_a} to substitute the left hand side
by
\[
\frac{1}{\ell}  \varphi=-\frac{\beta}{2\ell a}\pi_a=\frac{\beta}{\ell\gamma}\frac{1}{\lapseN}\frac{d a}{dt}.
\]
As for the right hand side, if we denote by $\scalef$ the scale factor of RW, then $\Theta=3 U(\scalef)/\scalef$,
and therefore \eqref{varphi_expansion} reduces to
\begin{align}\label{expansions_l_g_pre}
  \frac{\beta}{\gamma\ell} \frac{1}{\lapseN}\frac{d a}{dt}= \frac{U(\scalef)}{\scalef},
\end{align}
which, using the correspondence $\lapseN^{-1}\frac{d}{dt}=U$,
can be written as
\begin{align}\label{expansions_l_g}
  \frac{\beta}{\gamma\ell} U(a)= \frac{U(\scalef)}{\scalef}.
\end{align}
This relation establishes a direct correspondence between
$a$ and $\scalef$, the scale factor of RW, if 
\begin{equation}\label{l_g}
  \frac{\beta}{\gamma}=\frac{\ell}{a}.
\end{equation}
This corresponds to the relation found in \cite{Cendal:2024uzu}
using the interpretation of the twist as the extrinic curvature in \cite{freidelTwistedGeometriesGeometric2010a}.
At this point this choice may seem ad-hoc, but we are going to see next that
the second consistency check indeed yields
the direct correspondence between $a$ and the RW scale factor,
and thus \eqref{l_g} follows indeed for consistency.
 
For the second consistecy check we use the relation of the fluxes $X$ with the densitized vectors $E$ of the classical loop gravity description of the intrinsic geometry of  ``space'' $\Sigma$ in the Hamiltonian formulation of GR. The pass to the continuum limit 
(see e.g. \cite{freidelTwistedGeometriesGeometric2010a}),
is usually called to allow establishing that $X_i\simeq E_i$ for each  link $i$.
This relation may allow us to translate the evolution of $\vec X_i$ from \eqref{frame evolution}
to (dividing by $\lapseN$ at both sides)
\begin{align}\label{evolution of E}
[U,E_i]=\frac{2}{3}\frac{1}{\lapseN}\expansio E_i =  \frac{2}{3} \expansion E_i,
\end{align}
where here we also denote by $E_i$ the lift to $M$ of the vectors $E_i$ on $\Sigma$
($E_i$ are now vector fields on $M$ tangent to the foliation),
and for the second equality we have used the proper expansion in the two-vertex model
\[
  \expansion:=\frac{1}{\lapseN}\expansio,
\]
that describes the evolution of the volume with respect to the evolution operator $\lapseN^{-1}d/dt$.

Consider now one densitized vector $E_i$. We can choose
a triad (orthonormal) basis of vectors\footnote{Let us recall that the subindices $I$ in the triad
    enumerate the vectors of the basis, but also correspond eventually to the internal  $\alsu(2)$-valued components of the cotriad (or superindices in $\mathbb{R}^3$).}  $\{e_I\}$
so that one of them, say $\onee_i$, is aligned with $E_i$, and thus
$$E_i= \volg \onee_i,$$
where we use the shorthand $\volg:=\sqrt{\det \gammam_{ab}}$
with the metric writen in the basis $\gammam_{ab}=e^I_a e^J_b\delta_{IJ}$
where the set of forms $\{e^I\}$ is the dual basis of $\{e_I\}$. 
Then, \eqref{evolution of E} yields
\[
  [U,\onee_i]=\left(\frac{2}{3}\expansion -\frac{U(\volg)}{\volg}\right) \onee_i.
\]
Now, since $\onee_i$ is unit, then it holds
\[
  0=U(\gammam(\onee_i, \onee_i))=(\Lie_U \gammam)(\onee_i,\onee_i)+2\gammam([U,\onee_i],\onee_i),
\]
and therefore
\begin{equation}\label{Lie U one}
  (\Lie_U \gammam)(\onee_i,\onee_i)=-2\left(\frac{2}{3} \expansion -\expansionrw\right) \gammam(\onee_i,\onee_i)
\end{equation}
after using $\expansionrw=U(\volg)/\volg$.
On the other hand, using that $K=\frac{1}{2}\Lie_U\gammam$ ($U$ is normal to
the foliation $\{\Sigma_t\}$ and unit),
point (ii) above (for all $t$) is equivalent to
\begin{equation}\label{Lie U two}
  \Lie_U \gammam=\frac{2}{3}\Theta \gammam.
\end{equation}
The combination of \eqref{Lie U one} and \eqref{Lie U two} yields
\begin{align}\label{equal expansions}
 \expansion=\expansionrw,
\end{align}
that is, the expansion of the polyhedra (taking into account the lapse)
equals the expansion of RW
in this interpretation, as expected.
Clearly, the equality of the expansions provides the correspondence between
$a$ and the RW scale factor $\scalef$
once we observe that, by \eqref{can_transf} and \eqref{V evolution},
equation \eqref{equal expansions}
is equivalent to
\begin{equation}\label{equal a}
  \frac{1}{\lapseN} \frac{1}{a}\frac{da}{dt}=\frac{U(\scalef)}{\scalef}.
  \end{equation}
From the equations \eqref{equal a}  and \eqref{expansions_l_g_pre} we obtain \eqref{l_g},
as anticipated.

\section{Conclusions}

The frame bases described in section \ref{Sec_Frame_basis} have been shown to provide a geometrical interpretation of the twisted geometries variables on a graph, solely based on $\mathbb{R}^{3}$ vectors and spatial rotations. Using that representation, we have achieved a full description of the time evolution  of twisted geometries in the two-vertex model with the dynamics generated by a generalized Hamiltonian \eqref{Hamiltonian_equation2}. This general result has been particularized to the $\U(N)$-reduced sector of the model and we have proved that, remarkably, the planar angles of the faces of polyhedra remain constant in time, implying an homothetic (isotropic) expansion of the polyhedra, as it had been suggested in \cite{Cendal:2024uzu}.

  As a consequence of the homothetic expansion of polyhedra in the $\U(N)$ sector of the model, we have been able to complement the results found in \cite{Cendal:2024uzu} to ultimate the correspondence between the RW geometry and that described by the twisted geometries in this reduced sector of the model. In particular, the characterization of RW in terms of the conditions (i)-(iv) presented in section \ref{Sec_cosmo} has been explicitly identified with (i) the homogeneity and isotropy of the twisted geometries on the two-vertex graph, (ii)-(iii) the fact that all twist angles are independent of the link of the graph and (iv) the fact that the lapse function only depends on time. Moreover, we have shown that the time evolution of the normal vectors to the faces of polyhedra in the $\U(N)$-reduced sector of the two-vertex model coincides with the time evolution of the densitized triads $E_i$ in a RW geometry, as well as the expansion of polyhedra $\theta$ coincides with the expansion $\Theta$ in the RW geometry.

\acknowledgments
We would like to thank Álvaro Cendal, Luis J. Garay and Diego H. Gugliotta for very enlightening conversations. This work is supported by the Basque Government Grant IT1628-22, and by the Grant PID2021-123226NBI00 (funded by MCIN/AEI/10.13039/501100011033 and by ``ERDF A way of making Europe''). Additionally, S. R. acknowledges financial support from MIU (Ministerio de Universidades, Spain) fellowship FPU23/01491.

\bibliography{bibliography}

\end{document}